\documentclass[12pt]{article}
\usepackage{amsmath}
\usepackage{graphicx,psfrag} 
\usepackage{enumerate}
\usepackage{natbib}

\newcommand{\nuc}{\newcommand}
\nuc{\cc}{\cite}
\nuc{\bul}{\bullet}

\nuc{\wtN}{1,\cdots,N}
\nuc{\ben}{\begin{enumerate}}
\nuc{\een}{\end{enumerate}}

\nuc{\Ra}{\Rightarrow}
\newcommand{\EB}{\begin{eqnarray*}}
\newcommand{\EE}{\end{eqnarray*}}

\nuc{\tht}{\theta}
\nuc{\EQ}{\begin{displaymath}}
\nuc{\EN}{\end{displaymath}}

\newcommand{\blind}{0}

\begin{document}



%
%
%
%
%
%
%




\def\spacingset#1{\renewcommand{\baselinestretch}%
{#1}\small\normalsize} \spacingset{1}


\if0\blind
{
  \title{\bf Pearson Distance is not a Distance}
  \author{Victor Solo\hspace{.2cm}\\
    Department of Electrical Engineering, University of New South Wales
}
  \maketitle
} \fi

\if1\blind
{
  \bigskip
  \bigskip
  \bigskip
  \begin{center}
    {\LARGE\bf Pearson Distance is not a Distance}
\end{center}
  \medskip
} \fi

\bigskip
\begin{abstract}
The Pearson distance between a pair of random variables $X,Y$ with correlation $\rho_{xy}$,
namely, 1-$\rho_{xy}$, has gained widespread use,
particularly for clustering, in areas such as
gene expression analysis, brain imaging and cyber security. 
In all these applications it is implicitly assumed/required that
the distance measures be metrics, thus satisfying
the triangle inequality.
We show however, that
Pearson distance is not a metric. We go on to
show that this can be repaired by recalling the result,  
(well known in other literature) that $\sqrt{1-\rho_{xy}}$ is a metric.
We similarly show that a related measure of interest,
$1-|\rho_{xy}|$, which is invariant to the sign of $\rho_{xy}$, is not
a metric but that $\sqrt{1-\rho_{xy}^2}$ is.
We also give generalizations of these results.
\end{abstract}

\noindent%
\vfill

\newpage
\spacingset{1.45} 

\nuc{\disy}{dissimilarity }
\nuc{\disY}{dissimilarity}




\nuc{\rand}{random }
\nuc{\simy}{similarity }
\nuc{\semd}{semi-definite }
\nuc{\semD}{semi-definite}

\nuc{\varib}{variable }
\nuc{\varibs}{variables }
\nuc{\varibS}{variables}

\nuc{\aij}{a_{ij}}
\nuc{\aii}{a_{ii}}

\nuc{\dxy}{d(X;Y)}
\nuc{\rhoxy}{\rho_{x,y}}
\nuc{\csq}{cos^2(\tht)}
\nuc{\cs}{cos(\tht)}
\nuc{\csf}{cos^4(\tht)}

\nuc{\dij}{d_{ij}}
\nuc{\dik}{d_{ik}}
\nuc{\dkj}{d_{kj}}

\nuc{\dnxy}{d_{xy}}
\nuc{\dnxz}{d_{xz}}
\nuc{\dnzy}{d_{zy}}
\nuc{\dnyz}{d_{yz}}



\nuc{\pep}{\frac{1}{1+\ep}}
\nuc{\piot}{\frac{\pi}{2}}

\nuc{\sij}{s_{ij}}
\nuc{\sii}{s_{ii}}
\nuc{\sgsqu}{\sg^2_U}
\nuc{\sgu}{\sg_U}
\nuc{\rhoab}[2]{\rho_{#1,#2}}
\nuc{\rhoba}[1]{\rho_{#1}}

\nuc{\ssq}{sin^2(\tht)}
\nuc{\ssqh}{sin^2(\frac{\tht}{2})}

\nuc{\rij}{\rho_{ij}}
\nuc{\ups}{^{\sqrt{}}}
\nuc{\rxy}{\rho_{xy}}


\nuc{\wpep}{\frac{1+\ep}{2}}

\label{sec:intro}
Distance measures have a long history of application in Statistics as well 
as in many other areas.
Some of the earliest 
applications in Statistics are to 
cluster analysis, \cc{GOWJ67},\cc{SOKS63} and
multidimensional scaling, \cc{TORG58},\cc{KRUS77},\cc{GOWJ66}.
More recently there
 has been an explosion of interest 
in distance measures
 for application to
gene expression data, \cc{GIBR02},\cc{HAES05},\cc{JASK14}.
Other recent sources of interest include brain imaging \cc{KREG16} and
cyber security, \cc{SODE15}.

Note that we are concerned here with
measures of distance between \rand \varibS,
not with measures of distance between probability
measures - such as Hellinger distance.
To prevent confusion between 'distance' and 'metric', henceforth
we will instead use the pair 'dissimilarity' and 'metric', following
 terminology of \cc{GOWJ86}.

$\bul$ Given a finite set of \rand \varibs $X_k,k\wtN{1}{N}$, 
a pairwise dissimilarity between $X_i,X_j$ 
 is a functional of the bivariate distribution
satisfying:
(i) symmetry: $\dij=d_{ji}$;
(ii) strict positivity i.e. $\dij\geq 0$ with equality iff $i=j$.
A normalised pairwise similarity measure is a functional of the bivariate distribution
which satisifies $|\sij|\leq 1$, $\sii=1$ and symmetry.
Associated to any normalised similarity measure is a dissimilarity measure $\dij=1-\sij$.

$\bul$ A dissimilarity measure $\dij$ is a metric if it obeys the
triangle inequality: $\dij\leq\dik+\dkj$ for all $1\leq i,j,k\leq N$.

But why do we need the metric property? Most of the applications above are concerned
with clustering which needs to be done consistently. To partition variables into two or
more clusters coherently, we need to ensure that the variables in one cluster are closer to
each other than they are to the variables in another cluster. If X is close to Y (so that $\dnxy$
is small) and Z is close to Y (so that $\dnzy$ is small) then to put X,Y,Z in the same cluster
we need to know that X is close to Z; so we need to ensure $\dnxz$ is small. The simplest
way to guarantee this is to require the triangle inequality holds, otherwise we can have $\dnxz$
arbitrarily large. Thus clustering is made coherent by ensuring $\dij$ is a metric
\footnote{An even stronger condition, not pursued here, is the ultrametric
condition $\dnxz\leq max(\dnxy,\dnyz)$.}


Continuing, clearly the pairwise correlation $\rij$ is a normalised similarity measure
and we will focus on similarity/dissimilarity measures that are functions of correlation. Of
particular interest will be four measures of dissimilarity:
\ben
\item Pearson dissimilarity: $1-\rij$.
\item $|P|$earson dissimilarity: $1-|\rij|$.
\item $\ups$Pearson dissimilarity: $\sqrt{1-\rij}$.
\item P$^2$earson dissimilarity: $\sqrt{1-\rij^2}$.
\een
Numerous dissimilarity measures have been
applied to gene expression data but 
Pearson dissimilarity is 
one of the main ones:
\cc{GIBR02}[definition on p1575],
\cc{HAES05}[Table 1],
\cc{JASK14}[comments below equation (11)],
\cc{SODE15}[equation (41)].
The story is similar in brain imaging,
\cc{KREG16}[section 2.2] and cyber security, \cc{SODE15}[equation (41)].

\cc{GOWJ66} discussed normalised similarity informally,
showing that $\sqrt{1-\sij}$ is a metric
if the matrix $[\sij]$ is positive semi-definite.
These results are formally proved in \cc{GOWJ86}.
While \cc{GOWJ66} does not mention correlation explicitly
we may say that it has been known for at least half a century
that $\ups$Pearson dissimilarity, is a metric.
But what of Pearson dissimilarity?

In section II 
we use elementary arguments to
show that Pearson dissimilarity is not a metric
and also provide an elementary proof
that $\ups$Pearson \disy is a metric.
In section III we use similar elementary arguments to show that
$|P|$earson dissimilarity is not a metric, while P$^2$earson \disY, is.
In section IV contains conclusions.
For completeness, in the appendix, we discuss more general
results from the metric preserving literature and the work of \cc{GOWJ86}.
\section{Pearson Dissimilarity}

We give elementary analyses of
$\ups$Pearson \disy and
Pearson \disY.


{\bf Result I}.  $\ups$Pearson \disy  is a Metric.\\
{\it Proof}.
Let $(X,Y,Z)=(X_1,X_2,X_3)$ be zero mean random variables each with unit variance.
Then consider that
\EB
|X-Y|&\leq& 
|X-Z|+|Z-Y|\\
\Ra (X-Y)^2&\leq&
 (X-Z)^2+(Z-Y)^2+2|X-Z||Z-Y|\\
\Ra E(X-Y)^2&\leq &
E(X-Z)^2+E(Z-Y)^2
+2E(|X-Z||Z-Y|)\\
&\leq& 
E(X-Z)^2+E(Z-Y)^2
+2\sqrt{E(X-Z)^2}\sqrt{E(Z-Y)^2}\\
&=&
[\sqrt{E(X-Z)^2}+\sqrt{E(Z-Y)^2}]^2\\
\Ra
\sqrt{E(X-Y)^2}&\leq& \sqrt{E(X-Z)^2}+\sqrt{E(Z-Y)^2}\\
\Ra \sqrt{2(1-\rhoba{x,y})}&\leq&
\sqrt{2(1-\rhoba{x,z})}+\sqrt{2(1-\rhoba{z,y})}.
\EE
and the proof is complete.\\

{\bf Result II}.  Pearson \disy is not a Metric.\\
{\it Proof}.
Let $U,V,W$ be zero mean
random variables with:\\
$U$ independent of $V,W$ and for $0<\tht<\frac{\pi}{2}$
\EQ
var(V)=var(W)=\ssq,corr(V,W)=-\csq\mbox{ and }var(U)=\csq.
\EN
Then set
$(X,Y,Z)=(U+V,U+W,U)$ so that
$var(X)=1=var(Y)$. Thus
\EQ
\rhoab{y}{z}=\rhoab{x}{z}=\frac{E(XZ)}{\sqrt{var(X)var(Z)}}=\frac{\csq}{|\cs|}=|\cs|.
\EN
However $|\cs|=\cs$ since $0<\tht<\piot$. Continuing
\EB
\rhoab{x}{y}&=&E(U+V)(U+W)=E(U^2)+E(VW)\\
&=&var(U)+cov(V,W)\\
&=&var(U)+corr(V,W)\sqrt{var(V)var(W)}\\
&=&\csq-\csq\ssq=\csf
\EE
Now we show the triangle inequality can be violated i.e.
\EB
(1-\rhoba{x,y})>
(1-\rhoba{x,z})&+&(1-\rhoba{y,z})=2(1-\rhoba{x,z})\\
\equiv 1-\csf&>&2(1-\cs)\\
\equiv (1-\csq)(1+\csq)&>&2(1-\cs)\\
\equiv (1+\cs)(1+\csq)&>&2
\EE
This certainly holds for $|\tht|\leq \frac{\pi}{4}$
and the claim is established. Not that the violation is not infinitesimal,
it is gross, covering over half the range of $\tht$ values.

\section{P$^2$earson and $|P|$earson Dissimilarities}
Here we give elementary analyses of P$^2$earson and $|P|$earson \disY.

{\bf Result III}. P$^2$earson \disy is a metric.\\

{\it Proof}. This result is due to \cc{MATE08} but
we give an elementary proof for completeness.
Let $X_k,k\wtN{1}{N}$ and $Y_k,k\wtN{1}{N}$
be two independent
random vectors each with zero mean, unit variances  
and the same correlation matrix $[\rij]$. Set $Z_k=X_kY_k,k\wtN{1}{N}$.
Then the $Z_k$ have zero means and unit variances since
\EB
&&E(Z_k)=E(X_kY_k)=E(X_k)E(Y_k)=0\\
&&var(Z_k)=E(Z^2_k)=E(X_k^2Y_k^2)=E(X_k^2)E(Y_k^2)=1.
\EE
Their correlations are, for $i\neq j$
\EQ
E(Z_iZ_j)=E(X_iY_iX_jY_j)=E(X_iY_i)E(X_jY_j)=\rij^2
\EN
Now result I applies to the set of \rand \varibs $\{Z_k\}$ and the result follows.\\

{\bf Result IV}.  $|P|$earson \disy  is not a metric.

{\it Proof}. We use the construction from Theorem 2.2.
We show the triangle inequality is violated. We need
\EB
1-|\rhoba{x,y}|&>&1-|\rhoba{x,z}|+1-|\rhoba{y,z}|\\
\equiv 1-cos^4(\tht)&>&2-2cos(\tht)\\
\equiv 2cos(\tht)&>&1+cos^4(\tht)\\
\equiv cos(\tht)(2-cos^3(\tht))&>& 1
\EE
One can easily check that this holds for $\frac{\pi}{6}\leq \tht\leq \frac{\pi}{4}$;
again a gross violation.

\section{Conclusion}

In this paper, motivated by the widespread use of distance measures for 
clustering in numerous applications we explained why coherent clustering 
necessitates the metric property.
We have then given elementary analyses of four distance 
measures followed by an appendix
which sketches more general results.

The most widely used measure, Pearson distance $=1-\rxy$, is not a metric whereas
it has long been known, that a simple modification, $\ups$Pearson $=\sqrt{1-\rxy}$ is a metric. A
popular sign invariant modification of Pearson distance,  $|P|$earson = $1-|\rxy|$ is also not a
metric whereas a simple modification P$^2$earson = $\sqrt{1-\rxy^2}$ is a metric.

\section{Appendix: General Methods}
The results of \cc{GOWJ86} provide a general
approach to proving the metric property 
of a \disy measure associated with a normalised similarity measure.
And methods from the metric preserving function literature, \cc{CORA99}
provide general tools for proving when transformations of metrics
succeed or fail to preserve the metric property.
We state some of the basic results and then  apply them to the current setting.

{\bf Result S1}. \cc{GOWJ86}. If $[\sij]$ is a normalised similarity matrix
then the \disy $\dij=\sqrt{1-\sij}$ is a metric if $[\sij]$ is positive semi-definite.

This immediately implies result I.

To continue we introduce:\\
$\bul$ A function $f(x)$ is
metric preserving if whenever $\dij$ is a metric so is $f(\dij)$.

Note that we need only consider $f(x)$ for $x\geq 0$ since metrics are non-negative.

We now have the following results.\\

{\bf Result M1}.  \cc{CORA99}. If g(x) is strictly convex and passes through the origin
i.e. g(0) = 0 then it is not metric preserving.

{\bf Result M2}. \cc{CORA99}. If f(0) = 0 and f(x) is concave and strictly increasing
then it is metric preserving.

From result M1 we immediately get:

{\bf Result II$^\ast$}.  
No strictly convex function of (a) $\sqrt{1-\rij}$
or of (b) $\sqrt{1-\rij^2}$, which passes through the origin,
 is a metric.

Result II follows from
result II$^\ast$a if we take $g(x)=x^2$.
Result IV follows from result II$^\ast$b  as follows.
Set 
\EQ
\dij=\sqrt{1-\rij^2}
\Ra 1-|\rij|=1-\sqrt{1-\dij^2}=g(\dij)
\EN
and note that $g(d)=0$ iff $d=0$. Next
\EQ
g'(d)=\frac{d}{\sqrt{1-d^2}}
\Ra g''(d)=\frac{1}{\sqrt{1-d^2}(1-d^2)}>0
\EN
so that $g(d)$ is strictly convex and result IV follows.

Result M2 immediately delivers:

{\bf Result IV$^ast$}. Any strictly increasing concave function of
$\sqrt{1-\rij}$ or of $\sqrt{1-\rij^2}$, that passes through the origin, is a metric.

{\it Example 4.1}. 
Since $f(x)=\sqrt{x}$ is concave then $\sqrt{\dij}=(1-\rij)^{1/4}$ is a metric.\\


\end{document}